\documentclass[12pt]{iopart}

\usepackage{iopams}
\usepackage{color}
\usepackage{graphicx}
\usepackage{epsf, subfigure, verbatim}
\usepackage{psfrag}
\usepackage{amsmath2}
\usepackage{cite}
\newcommand{\ben}{\begin{enumerate}}
\newcommand{\een}{\end{enumerate}}
\newcommand{\be}{\begin{equation}}
\newcommand{\ee}{\end{equation}}
\newcommand{\bse}{\begin{subequation}}
\newcommand{\ese}{\end{subequation}}
\newcommand{\bea}{\begin{eqnarray}}
\newcommand{\eea}{\end{eqnarray}}
\newcommand{\bc}{\begin{center}}
\newcommand{\ec}{\end{center}}

\begin{document}

\title[MCS]{Links between symmetry reduction and Hirota methods of the $\mathcal{N}=2$
susy KdV equation}

\author{L. Delisle and V. Hussin$^1$}
\address{$^1$D\'epartement de Math\'ematiques et de 
Statistique et Centre de Recherches Math\'{e}matiques,
Universit\'{e} de Montr\'{e}al, Montr\'{e}al, Qu\'ebec, H3C 3J7, Canada} 
\ead{delisle@DMS.UMontreal.ca;
hussin@DMS.UMontreal.ca}
\begin{abstract}
We consider the resolution of the $\mathcal{N}=2$ supersymmetric KdV equation with $a=-2$ ($SKdV_{a=-2}$) from two approaches, the group invariant method (or symmetry reduction) and the Hirota formalism. A bilinear form of the $SKdV_{a=-2}$ equation is constructed. Links between the two methods are established and new solutions are obtained from both approaches.

\bigskip


\bigskip

\end{abstract}

\maketitle \setcounter{equation}{0}
\section{Introduction \label{sec:1}}
Let us recall that $\mathcal{N}=2$ supersymmetric (SUSY) extensions of the KdV equation have been largely studied in the past \cite{LM,AHW,IMM,GS,HK} in terms of integrability conditions and solutions. Such extensions are given as a one-parameter ($a\in \mathbb{R}$) family of Grassmann-valued partial differential equations with one dependent variable $A(x, t;\theta_{1}, \theta_{2})$ where the independent variables are given as a set of even (commuting) space $x$ and time $t$ variables and a set of odd (anticommuting) variables $\theta_{1}$, $\theta_{2}$. The dependent variable $A$ is supposed to be an even superfield and satisfies 
\begin{equation}
 A_{t}=(-A_{xx}+(a+2) AD_{1}D_{2}A+(a-1)(D_{1}A)(D_{2}A)+a A^{3})_{x},
 \label{N2mkdv}
\end{equation}
 where $D_{1}$, $D_{2}$ are the covariant superderivatives given by
\begin{equation}
 D_{i}=\theta_{i}\partial_{x}+\partial_{\theta_{i}},\quad i=1,2,
\end{equation}
and such that $D_{1}^2=D_{2}^2=\partial_{x}$.

Since the odd variables satisfy $\theta_{1}^2=\theta_{2}^2=\theta_{1}\theta_{2}+\theta_{2}\theta_{1}=0$, the dependent variable $A$ admits the following Taylor expansion 
\begin{equation}
 A(x,t;\theta_{1},\theta_{2})=u(x,t)+\theta_{1}\xi_{1}(x,t)+\theta_{2}\xi_{2}(x,t)+\theta_{1}\theta_{2}v(x,t),
 \label{decompA}
\end{equation}
where $u$ and $v$ are commuting (bosonic) complex valued functions and $\xi_{1}$ and  $\xi_{2}$ are anticommuting (fermionic) complex valued functions.
Equation (\ref{N2mkdv}) can thus be decomposed as a set of two bosonic equations 
\begin{align}
u_{t}+(u_{xx}-au^3+(a+2)u v-(a-1)\xi_{1}\xi_{2})_x&=0,\label{GetmKdV}\\
v_{t}+(v_{xx}+3v^2-3 a u^2v-(a+2)u u_{xx}
-(a-1) u_{x}^2)_x&\phantom{=}\notag\\
-(3\xi_{2}\xi_{2; x}+(a+2)\xi_{1}\xi_{1; x}-6a u \xi_{1}\xi_{2})_x&=0\label{GetKdV}
\end{align}
and two fermionic ones
 \begin{eqnarray}
&\xi_{1; t}+(\xi_{1; xx} +3(v-a u^2)\xi_{1}- (a+2) u \xi_{2; x}-(a-1) u_{x}  \xi_{2} )_x=0,
\label{ksi1}\\
&\xi_{2; t}+(\xi_{2; xx} +3(v-a u^2)\xi_{2}+(a+2) u \xi_{1; x}+(a-1) u_{x}  \xi_{1} )_x=0.
\label{ksi2}
\end{eqnarray}
We easily see that this last set can be decoupled. Indeed, taking $\rho_{\pm}=\xi_{1} \pm i \xi_{2}$, we get
 \begin{equation}
\rho_{\pm; t}+(\rho_{\pm; xx} +3(v-a u^2)\rho_{\pm}\pm i(a+2) u \rho_{\pm; x}\pm i(a-1) u_{x}  \rho_{\pm} )_x=0.
\label{rhopm}
\end{equation}

Solutions of equation (\ref{N2mkdv}) have been obtained using different methods. Among them, an original approach \cite{AHW} has adapted the classical symmetry reduction method to the SUSY context. Indeed in this approach, starting with the SUSY equation (\ref{N2mkdv}), it has been reduced to a set of only one bosonic and one fermionic equations using invariance properties associated with invariant superalgebras. 

It looks like a reduction from $\mathcal{N}=2$ to $\mathcal{N}=1$ SUSY  but it is not quite the same as we will see later. It also gives different results than a reduction method which consists of setting to zero some of the dependent variables (see \cite{HK} for example).

Let us mention that the bosonization approach \cite{ARS} has also been used to solve such systems. It consists in expanding the bosonic and fermionic fields in $n$-fermionic parameter space. In the following developments, we will consider the simplest case (one-fermionic parameter bosonization) which consists of writing the fermionic fields as  $\xi_i={\zeta} f_i (i=1,2)$, with $f_i$ a bosonic complex function and $\zeta$ an odd parameter such that ${\zeta}^2=0$. In the decomposed original system of equations (\ref{GetmKdV}) and (\ref{GetKdV}), the fermionic contributions thus disappear and we get purely bosonic equations: 
\begin{align}
u_{t}+(u_{xx}-au^3+(a+2)u v)_{x}&=0,
\label{GetmKdVred}\\
v_{t}+(v_{xx}+3v^2-(a+2)u u_{xx}
-(a-1) u_{x}^2-3 a u^2v)_x&=0.\label{GetKdVred}
\end{align}
It is interesting to see that for $v=\pm i u_{x}$, (\ref{GetKdVred}) is a consequence of (\ref{GetmKdVred}).

 This is not the case for more general bosonization procedure. For example, in the 2-fermionic parameter space spanned by $\zeta_{1}$ and $\zeta_{2}$, the bosonic fields $u$ and $v$ takes the forms $u_{0}+\zeta_{1}\zeta_{2}u_{12}$ and $v_{0}+\zeta_{1}\zeta_{2}v_{12}$, respectively. We see that the functions $u_{0}$ and $v_{0}$ satisfy the same set of equations (\ref{GetmKdVred}) and (\ref{GetKdVred}). On the other hand, the fermionic fields $\xi_{1}$ and $\xi_{2}$ have the new forms $\zeta_{1}p_{1}+\zeta_{2}p_{2}$ and $\zeta_{1}q_{1}+\zeta_{2}q_{2}$, so that the set of fermionic equations (\ref{ksi1}) and (\ref{ksi2}) is doubled. The evolution equations on $u_{12}$ and $v_{12}$ involve all the bosonic and fermionic dependent variables.

In the following, we will take $a=-2$ in equation (\ref{N2mkdv}) where it is well-known that we get solitons solutions as travelling wave solutions \cite{AHW,IMM,GS} and also rational similarity solutions as in the classical case \cite{K,DJ}. Such solutions will be generalized in the SUSY case using the symmetry reduction method and also the Hirota formalism.  

Let us write the set of equations we are working with in this case. The equation (\ref{N2mkdv}) becomes: 
\begin{equation}
 A_{t}=(-A_{xx}-3(D_{1}A) (D_{2}A)-2 A^{3})_{x},
 \label{N2mkdva-2}
\end{equation}
and admits the decomposition
 \begin{align}
u_{t}+(u_{xx}+2 u^3+3 \xi_{1}\xi_{2})_x&=0,\label{GetmKdVa-2}\\
v_{t}+(v_{xx}+3v^2+6 u^2v+3 u_{x}^2)_x-(3\xi_{2}\xi_{2; x}+12 u \xi_{1}\xi_{2})_x&=0,\label{GetKdVa-2}\\
\xi_{1; t}+(\xi_{1; xx} +3(v+2 u^2)\xi_{1}+3 u_{x}  \xi_{2} )_x&=0,\label{ksi1a-2}\\
\xi_{2; t}+(\xi_{2; xx} +3(v+2 u^2)\xi_{2}-3 u_{x}  \xi_{1} )_x&=0.\label{ksi2a-2}
\end{align}
The equations (\ref{rhopm}) become
 \begin{equation}
\rho_{\pm; t}+(\rho_{\pm; xx} +3(v+2 u^2 \mp i u_{x} ) \rho_{\pm} )_x=0,
\label{rhopma-2}
\end{equation}
and thus equations (\ref{GetmKdVa-2}) and (\ref{GetKdVa-2}) can be written as:
 \begin{align}
u_{t}+(u_{xx}+2 u^3+\frac{3}{2} i \rho_{+}\rho_{-})_x&=0,\label{GetmKdVa3}\\
v_{t}+(v_{xx}+3v^2+6 u^2v
+3 u_{x}^2)_x+(\frac{3}{4}(\rho_{-}-\rho_{+})(\rho_{-; x}-\rho_{+; x})-6 i u \rho_{+}\rho_{-})_x&=0.\label{GetKdVa3}
\end{align}
In particular, for one-fermionic parameter bosonization, we get the following system of nonlinear partial differential equations :
 \begin{align}
u_{t}+(u_{xx}+2 u^3)_x&=0,
\label{GetmKdVaeven}\\
v_{t}+(v_{xx}+3v^2+6 u^2v
+3 u_{x}^2)_x&=0.\label{GetKdVaeven}
\end{align}
We see that $u$ satisfies the mKdV equation while $v$ solves a modified KdV equation.

\section{Symmetry reduction method and solutions}

Let  us here briefly recall the symmetry reduction method as it has been adapted to the SUSY context \cite{AHW}.  The Lie superalgebra of symmetries of equation (\ref{N2mkdv}) is a $(3|2)$-dimensional superalgebra with three even generators associated with time and space translations and dilations and two odd generators. We get explicitly;
\begin{equation}
{\cal P}_{t}=\partial_{t},\  {\cal P}_{x}=\partial_{x}, \ {\cal D}=x \partial_{x}+3 t \partial_{t}+\frac12 ( \theta_1 \partial_{\theta_1}+\theta_2 \partial_{\theta_2})-A \partial_{A},
\end{equation}
\begin{equation}
{\cal Q}_{1}=\theta_1 \partial_{x}-\partial_{\theta_1}, \ {\cal Q}_{2}=\theta_2 \partial_{x}-\partial_{\theta_2}.
\end{equation}
The non zero structure relations are given by
\begin{align}
[{\cal P}_{t},{\cal D}] = 3{\cal P}_{t}, \  [{\cal P}_{x},{\cal D}]&= {\cal P}_{x}, \ [{\cal Q}_{i},{\cal D}]= \frac12 {\cal Q}_{i}, \ i=1,2,\\
\{{\cal Q}_{1},{\cal Q}_{1}\}&= \{{\cal Q}_{2},{\cal Q}_{2}\}=-2 {\cal P}_{x}.
\end{align}
Two distinct sub-superalgebras are considered and used to reduce the equation (\ref{N2mkdv}) to a set of nontrivial ordinary differential equations.  The first sub-superalgebra $G_1$ is generated by $\{{\cal P}_{t}- c {\cal P}_{x},\ {\cal Q}_{1}+i {\cal Q}_{2}\}$ ($c$ is a real parameter) and corresponds to translational invariance that will give rise to the so-called travelling wave solutions. The second sub-superalgebra $G_2$ is generated by $\{{\cal D},\ {\cal Q}_{1}+i {\cal Q}_{2}\}$ and corresponds to dilation invariance that will give rise to the so-called similarity solutions.

In order to perform the symmetry reduction, we have to find the invariant of the action of the corresponding subgroup ${\cal{G}}_1$ or ${\cal{G}}_2$ on the independent and dependent variables and rewrite the equation (\ref{N2mkdv})  in terms of these invariants. For the subalgebra $G_1$ and corresponding subgroup ${\cal{G}}_1$, we get the invariants \cite{AHW}
\begin{equation}
y=x+c t+i \theta_1 \theta_2, \ \theta=\theta_1+i\theta_2, \ A=A(y,\theta) \label{invatra}
\end{equation}
and the reduced equation (after integration with respect to $y$) is
\begin{eqnarray}
&A_{yy}-a A^3+i(a+2) A A_y +c A \nonumber\\
&-2i(a+2) \theta A A_{y\theta}-2i(a-1) \theta A_{\theta}A_y+ c_1+\theta k_1=0, 
\label{A reduce1}
\end{eqnarray}
where $c_1$ and $k_1$ are complex integration constants. It is not a $\mathcal{N}=1$ SUSY reduction since equation (\ref{A reduce1}) cannot be expressed in terms of the 
superderivative  $D_{\theta}= \theta \partial_y+ \partial_{\theta}$. 

Using the decomposition of the invariant superfield $A(y, \theta)=u(y)+\theta\rho(y)$, we get the following system of ordinary differential equations:
\begin{align}
u_{yy}-a u^3+i(a+2) u u_y +c u+ c_1&=0, \label{redu}\\
\rho_{yy}-i(a+2) u \rho_{y}+(c-3a u^2+i(4-a) u_y) \rho +k_1&=0. \label{redrho}
\end{align}
Identifying  $A(y, \theta)$ with $A(x, t; \theta_{1}, \theta_{2})$ given by equation (\ref{decompA}), for $y_0=x+ct$,  we get
\begin{equation}
u(x,t)=u(y_0), v(x,t)=i\frac{d u(y_0)}{dy_0}, \xi_1(x,t)=-i \xi_2(x,t)=\rho(y_0).\label{decompositionA}
\end{equation}
We see that the two fermionic functions reduce to one independent function $\rho$ since $\rho_+=0$ and $\rho_-=2 \xi_1=2\rho$. 

For the subalgebra $G_2$ and corresponding subgroup ${\cal{G}}_2$, we get the invariants
\begin{equation}
z=t^{-\frac13}(x+i \theta_1 \theta_2), \ \theta=\theta_1+i\theta_2, \ W=t^{\frac13}A(z,\theta) \label{dilainv}
\end{equation}
and the reduced equation (after integration with respect to $z$) is
\begin{eqnarray}
&W_{zz} -aW^3-\frac13 z W+i(a+2) W W_{z}\nonumber \\
&-2i(a+2) \theta W W_{z\theta}-2i(a-1) \theta W_{\theta}W_z+ c_2+\theta k_2=0\label{A reduce2},
\end{eqnarray}
where $c_2$ and $k_2$ are complex integration constants. 

Since the invariant superfield $W$ can be decomposed as $W(z, \theta)=w(z)+\theta\phi(z)$, we get, from (\ref{A reduce2}), the following system of ordinary differential equations:
\begin{align}
w_{zz}-a w^3+i(a+2) w w_z -\frac13 z w+ c_2&=0, \label{redw}\\
\phi_{zz}-i(a+2) w \phi_{z}+(-\frac13 z-3a w^2+i(4-a) w_z) \phi +k_2&=0.\label{redphi}
\end{align}
The only difference between the sets of reduced equations (\ref{redu}) and (\ref{redrho}) is that the constant parameter $c$ has been replaced by $-\frac13 z$ in (\ref{redw}) and (\ref{redphi}).

Identifying  $A(z, \theta)=t^{-\frac13}W(z, \theta)$ with $A(x, t; \theta_{1}, \theta_{2})$ given by equation (\ref{decompA}), for $z_0=t^{-\frac13} x$,  we get
\begin{equation}
u(x,t)=t^{-\frac13} w(z_0), v(x,t)=it^{-\frac23}\frac{d w(z_0)}{dz_0}, \xi_1(x,t)=-i \xi_2(x,t)=t^{-\frac13}\phi(z_0).\label{identdila}
\end{equation}

As we see both reductions apply to equation (\ref{N2mkdv}) independently of the parameter $a$. 

\subsection{Soliton travelling wave solution for $a=-2$}

From (\ref{redu}) and (\ref{redrho}), we get the decoupled system of equations as
\begin{align}
u_{yy}+2 u^3+c u+ c_1&=0,\label{redu2}\\
\rho_{yy}+(c+6 u^2+6i u_y) \rho +k_1&=0.\label{redrho2}
\end{align}
General solutions of this set have been given in \cite{AHW}. Here we give some particular solutions, we will start with to make the connection with the Hirota formalism.

The mKdV equation (\ref{redu2}) admits the particular soliton solution 
\begin{equation}
u(y)={\text{sech}}(y),\label{solitu}
\end{equation}
if we take $c_1=0$ and  $c=-1$. The fermionic equation (\ref{redrho2}) thus admits the particular solution ($k_1=0$ and $\zeta$ is an odd parameter)
\begin{equation}
\rho(y)=\zeta\rho_{0}(y)={\zeta}\ {\text{sech}}^3(y)
\big( \sinh y +\frac{i}{2}(1-\sinh^2 y)\big). \label{fermiyrho}
\end{equation}

Using (\ref{decompositionA}), the original solution $A(x, t;\theta_{1}, \theta_{2})$ is thus given by (\ref{decompA}) with 
\begin{align}
u(x,t)&={\text{sech}}(x-t), \ v(x,t)=- i\ {\sinh (x-t)}\ {\text{sech}}^2 (x-t),\label{Bsoliton}\\ 
\xi_1(x,t)=-i \xi_2(x,t)&={\zeta} \ {\text{sech}}^3 (x-t) (\sinh (x-t) +\frac{i}{2}(1-\sinh^2 (x-t)).
\label{Fsoliton}
\end{align}
We see that we have indeed $v=i u_{x}$. 

Let us observe that using this symmetry reduction, we have in fact solved our original problem (\ref{N2mkdva-2}) with the following constraints $\xi_{1}=-i\xi_{2}$ and $v=i u_{x}$. This means that we have solved the set of decomposed equations:
\begin{align}
 u_{t}+(u_{xx}+2u^{3})_{x}&=0,\label{decusymetry}\\
\xi_{1;t}+(\xi_{1;xx}+6(iu_{x}+u^{2})\xi_{1})_{x}&=0.\label{decfsymetry}
\end{align}

Figure 1 shows the behaviour of $u$ as given in (\ref{solitu}) and the real and imaginary parts of $\rho_{0}$ given in (\ref{fermiyrho}) as a function of $y_{0}=(x-t)$.

\begin{figure}[!h]
\begin{center}
\includegraphics[width=8cm]{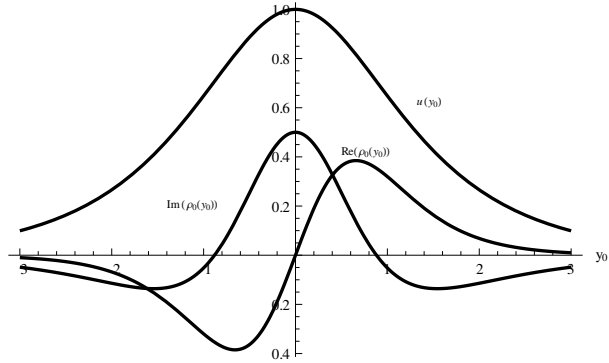}
\caption{The functions $u(y_{0})$, $Re(\rho_{0}(y_{0}))$ and $Im(\rho_{0}(y_{0}))$.}
\end{center}
\end{figure}

\subsection{Rational similarity solution for the case $a=-2$}
From (\ref{redw}) and (\ref{redphi}), we get the following system:
\begin{align}
w_{zz}+2 w^3-\frac13 z w+ c_2&=0 \label{redw2},\\
\phi_{zz}+(-\frac13 z+6 w^2+6 i w_z) \phi +k_2&=0.\label{redphi2}
\end{align}
The equation (\ref{redw2}) is known as the second  Painlev\'e equation and rational solutions for the bosonic mKdV equation are known \cite{K,DJ,FOU,AS,C}. A particular solution is (with $c_2=\frac23 i$):
\begin{equation}
w(z)=2i \frac{z^3-6}{z(z^3+12)}= i \frac{d}{dz}\log(\frac{z^3+12}{z}).
\label{wsim}
\end{equation}
With such a solution, the fermionic equation (\ref{redphi2}) becomes
\begin{equation}
\phi_{zz}-\frac{36 + z^3}{3 z^2} \phi =0,
\end{equation}
when $k_{2}=0$. Using the following change of variables 
\begin{equation}
 \phi\left(z\right)=\zeta\sqrt{z}\psi\left(\chi\left(z\right)\right), \quad \chi\left(z\right)= \frac{2}{3\sqrt{3}}z^{\frac{3}{2}},
\end{equation}
such that $\zeta^{2}=0$, we see that $\psi$ satisfies the modified Bessel equation,
\begin{equation}
 \chi^{2}\psi_{\chi\chi}+\chi\psi_{\chi}-\left(\chi^{2}+\nu^{2}\right) \psi=0,
\end{equation}
with $\nu=\pm\frac{7}{3}$. We get, in particular,
\begin{equation}
\phi\left(z\right)=\zeta\sqrt{z} I_{-\frac{7}{3}}\left(\frac{2}{3\sqrt{3}}z^{\frac{3}{2}}\right),\label{solphi}
\end{equation}
where $I_{\nu}(\chi)$ is a modified Bessel function. The original solution $A(x, t;\theta_{1}, \theta_{2})$ is thus given by equation (\ref{decompA}) where 
\begin{align}
u(x,t)=2i \frac{x^3-6t}{x(x^3+12t)}, v(x,t)&=2 \ \frac{x^6-48t x^3-72 t^2}{x^2(x^3+12t)^2},\label{Bsim}\\
\xi_1(x,t)=-i \xi_2(x,t)&=t^{-\frac13}\phi(t^{-\frac{1}{3}}x).
\end{align}
Here again, we see that $v=i u_{x}$. As before, $u$ and $\xi_1$ solve (\ref{decusymetry}) and (\ref{decfsymetry}) respectively.

Figure 2 shows the behaviour of the imaginary part of $u$ as a function of $x$ and $t$. Such solution of the ${\cal N}=2$ SUSY equation has not been considered before and may be compared with the two-soliton one that will be given in the next section using the Hirota formalism.

We refer to Appendix 1 for a discussion of the series of rational similarity solutions of equation (\ref{redw2}) constructed using the Yablonskii-Vorob'ev polynomials \cite{FOU,C} and the corresponding resolution of the fermionic equation (\ref{redphi2}).

\begin{figure}[!h]
\begin{center}
 \includegraphics[width=4.5cm]{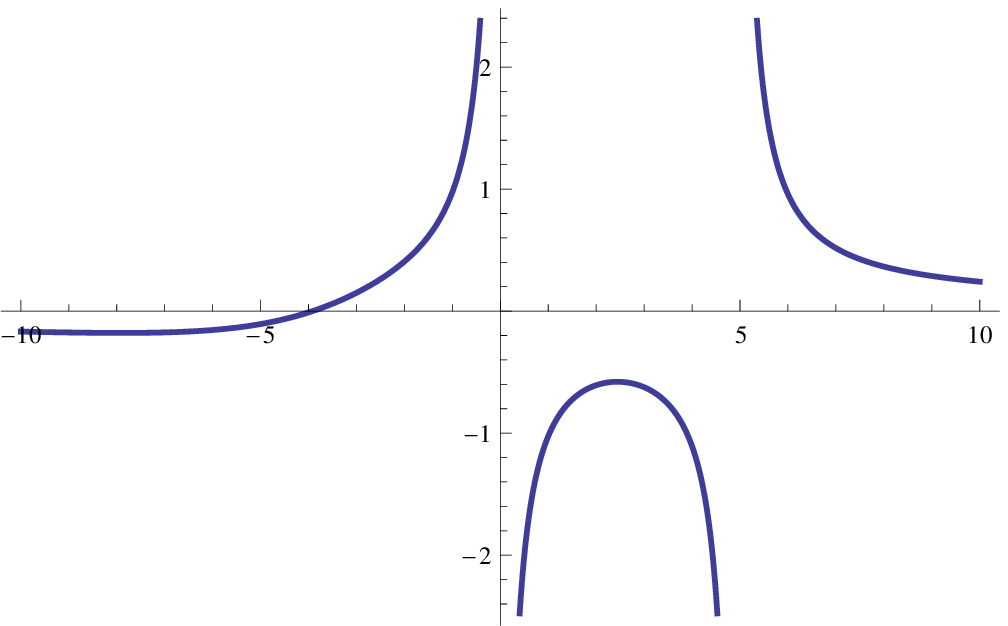}
\includegraphics[width=4.5cm]{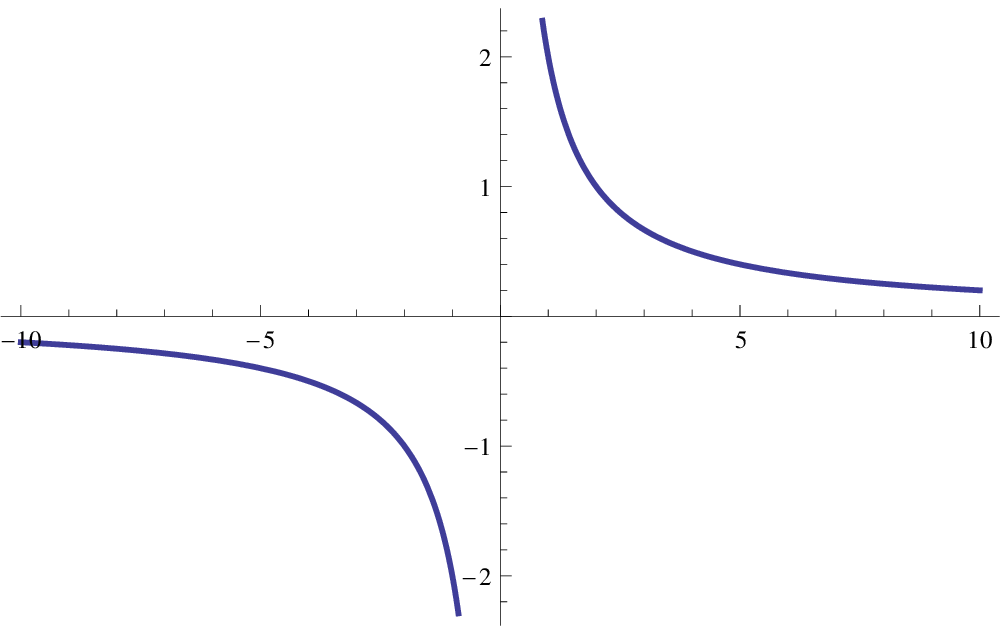}
\includegraphics[width=4.5cm]{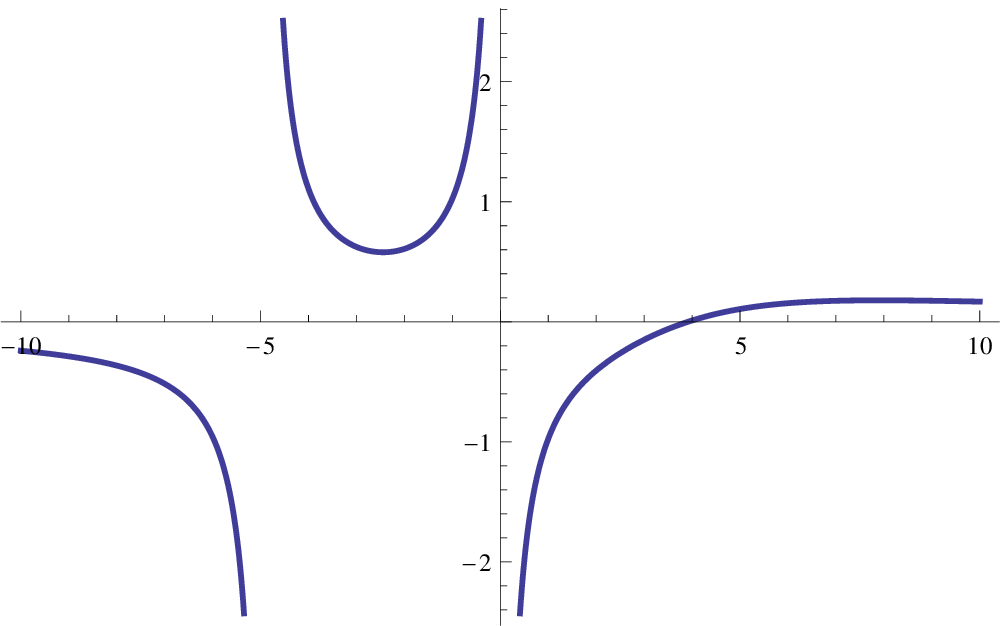}
\caption{The function $Im(u(x,t))$ for $t=-10, 0, 10$.}
\end{center}
\end{figure}



\section{Hirota formalism and solutions for the case $a=-2$}

The Hirota formalism is a well known process in the classical and in the $\mathcal{N}=1$  SUSY cases \cite{MY,C1,CRG,GS1}. This formalism has been used, in particular, to obtain $N$-soliton solutions. We examine the possible generalizations in the $\mathcal{N}=2$ SUSY case \cite{GS,HK} following the idea given in \cite{ZLW}.  We write the superfield (\ref{decompA}) in the following form
\begin{equation}
 A(x,t;\theta_{1},\theta_{2})=u^{b}(x,t;\theta_{1})+\theta_{2}\ \xi^{f}(x,t;\theta_{1}),
 \label{AHirota}
\end{equation}
 where $u^{b}, \ \xi^{f}$ are even and odd superfields, respectively. In fact, comparing with  (\ref{decompA}) , we have $u^{b}=u+\theta_{1}\xi_{1}$ and $\xi^{f}= \xi_{2}-\theta_{1} v$.We thus introduce the following change of variable $A=\partial_{x} B$, where 
\begin{equation}
B(x,t;\theta_{1},\theta_{2})={\cal U}^{b}(x,t;\theta_{1})+\theta_{2}\ \eta^{f}(x,t;\theta_{1}), 
\label{be}
\end{equation}
to equalize the order of the equation with the number of appearence of the $x$ derivative in the nonlinear terms. Note that since $D_{i}^{2}=\partial_{x}$ ($i=1,2$), $D_{i}$ can be viewed as the square root of $\partial_{x}$. Equation (11) thus becomes, after integrating once,
\begin{equation}
 B_{t}=-B_{xxx}-3(D_{1}B_{x})(D_{2}B_{x})-2B_{x}^{3},
 \label{eqube}
\end{equation}
where the constant of integration is set to zero. Inserting the explicit form (\ref{be}) of $B$ in equation (\ref{eqube}), we get a set of coupled $\mathcal{N}=1$ SUSY equations on ${\cal U}^{b}$ and $\eta^{f}$:
\begin{align}
 {\cal U}^{b}_{t}+{\cal U}^{b}_{xxx}-3\eta^{f}_{x}D_{1}{\cal U}^{b}_{x}+2({\cal U}^{b}_{x})^{3}&=0,
 \label{equv}\\
\eta^{f}_{t}+\eta^{f}_{xxx}-3\eta^{f}_{x}D_{1}\eta^{f}_{x}-3{\cal U}^{b}_{xx}D_{1}{\cal U}^{b}_{x}+6\eta^{f}_{x}
({\cal U}^{b}_{x})^{2}&=0.
\label{equeta}
\end{align}

 Now the strategy is to introduce a change of the dependent variables, ${\cal U}^{b}$ and $\eta^{f}$, in such a way that equations (\ref{equv}) and (\ref{equeta}) become quadratic. 
 



Since equation (\ref{equv})  is a modified mKdV equation, we use the change of variables:
\begin{equation}
 {\cal U}^{b}=\alpha \log(\dfrac{\tau_{1}}{\tau_{2}}), \quad \eta^{f}=\beta D_{1} \log(\dfrac{\tau_{1}}{\tau_{2}}), \label{veta}
\end{equation}
 where $\tau_{1}=\tau_{1}(x,t;\theta_{1})$ and $\tau_{2}=\tau_{2}(x,t;\theta_{1})$ are both bosonic superfields and the constants $\alpha$ and $\beta$ satisfy $\alpha^{2}=-1$ and $\beta^{2}=1$. With no lost of generality, we choose $\alpha=-i$ and $\beta=1$. 
 
Equation (\ref{veta}) represents  a constraint on the superfield $A$ since we get
\begin{equation}
 \eta^{f}=i D_{1}{\cal U}^{b}
 \label{etaspe}
\end{equation}
and thus
\begin{equation}
 \xi^{f}=i D_{1}u^{b}.
 \label{xispe}
\end{equation}
Using (\ref{xispe}) and comparing (\ref{decompA}) with (\ref{AHirota}), we get
\begin{equation}
 A(x,t;\theta_{1},\theta_{2})=u(x,t)+(\theta_{1}+i \theta_{2})\xi_{1}(x,t)-i \theta_{1}\theta_{2}u_{x}(x,t).
 \label{decompAspe}
\end{equation}
The constraints on the field $\xi_{2}$  and $v$ are very similar to the ones obtained in the symmetry reduction. We have already noticed that $v=\pm i u_{x}$ solves our problem when $u$ solves mKdV given in (\ref{GetmKdVaeven}). Here, we see that these constraints are imposed in order to produce a bilinear form.

Inserting equation (\ref{etaspe}) in equations (\ref{equv}) and (\ref{equeta}), we get
\begin{equation}
 {\cal U}^{b}_{t}+{\cal U}^{b}_{xxx}+2({\cal U}^{b}_{x})^{3}=0
 \label{mkdvv}
\end{equation}
and 
\begin{equation}
D_{1}\left({\cal U}^{b}_{t}+{\cal U}^{b}_{xxx}+2({\cal U}^{b}_{x})^{3}\right)=0.
\end{equation}
This last equation is trivially satisfied whenever ${\cal U}^{b}$ is a solution of the mKdV equation (\ref{mkdvv}). 
Let us mention that (\ref{mkdvv}) is still a SUSY equation and if $ {\cal U}^{b}$ takes the decomposition $ {\cal U}_0^{b}(x,t)+\theta_{1}\varphi(x,t)$, we get the classical mKdV equation for ${\cal U}_0^{b}$ and $\varphi$ satisfies
\begin{equation}
 \varphi_{t}+\varphi_{xxx}+6({\cal U}^{b}_{0; x})^{2}\varphi_{x}=0,
 \label{mkdvphi}
\end{equation}
for which a particular solution is $\varphi=\zeta \ {\cal U}^{b}_{0; x}$ where $\zeta$ is an odd constant. Such a result is common when we deal with ${\cal N}=1$ SUSY KdV and mKdV equations.

We thus have a direct bilinearization of equations  (\ref{equv}) and (\ref{equeta})
\begin{align}
 (\mathcal{D}_{t}+\mathcal{D}_{x}^{3})(\tau_{1}\cdotp\tau_{2})&=0,\label{H1}\\
\mathcal{S}\mathcal{D}_{x}(\tau_{1}\cdotp\tau_{2})&=0\label{H2},
\end{align}
where 
\begin{equation}
\mathcal{S}\mathcal{D}_{x}^{n}(\tau_{1}\cdotp\tau_{2})=(D_{\Theta_{1}}-D_{\Theta_{2}})(\partial_{x_{1}}-\partial_{x_{2}})^{n}
\tau_{1}(x_{1};\Theta_{1})\tau_{2}(x_{2};\Theta_{2})\lvert_{x=x_{1}=x_{2},\theta_{1}=\Theta_{1}=\Theta_{2}},
\end{equation}
is the super Hirota derivative and $D_{\Theta_{i}}=\partial_{\Theta_{i}}+\Theta_{i}\partial_{x_{i}}$, $i=1,2$. Equations (\ref{H1}) and (\ref{H2}) are a natural generalization to the susy case of the classical bilinear form of the mKdV equation.

Let us finally mention that, since we have 
\begin{equation}
u^{b}=u+\theta_{1}\xi_{1}=\partial_{x} {\cal U}^{b},
\end{equation}
such a bilinearization gives the following set of decomposed equations (together with $v=-i u_{x}$ and $\xi_{2}=i \xi_{1}$ ($\rho_{+}=0$)):
\begin{align}
u_{t}+(u_{xx}+2 u^3)_x&=0,\label{soluH}\\
\xi_{1; t}+(\xi_{1; xx} +6 u^2  \xi_{1})_{x} &=0, \label{solxiH}
\end{align}
which clearly solve (\ref{N2mkdva-2}) and such that a particular solution of the fermionic equation is $ \xi_{1}=\zeta u_{x}$. The comparison of this set of equations with the set (\ref{decusymetry}) and (\ref{decfsymetry}) obtained in the symmetry reduction, suggest that we will get different solutions for the fermionic components $\xi_{1}$ and $\xi_{2}$ of the superfield $A$.



\subsection{Super soliton solutions}

The Hirota formalism helps us to recover the travelling wave solution or one super soliton solution  but we also get the $N$ super soliton solutions \cite{GS,ZLW}.
Indeed, we take 
\begin{equation}
 \tau_{1}=1+a_{1} e^{\Psi}, \quad \tau_{2}=1+b_{1} e^{\Psi},
\end{equation}
where $\Psi=(\kappa x+\omega t)+\theta_{1}\zeta$ and $a_{1}$ and $b_{1}$ are nonzero even parameters. Introducing $\tau_{1}$ and $\tau_{2}$ in equation (\ref{H2}) yields the following relation
\begin{equation}
 b_{1}=-a_{1}.
\end{equation}
The dispersion relation
\begin{equation}
 \omega+\kappa^{3}=0,
\end{equation}
is obtained from the equation (\ref{H1}). Now, we have  $\Psi=\kappa (x-\kappa^2 t)+\theta_{1}\zeta$ which is similar to (\ref{invatra}) but not identical. It is expected since we are interested in travelling wave solutions.  Since 
\begin{equation}
 {\cal U}^{b}=F(\Psi)= F(\Psi_0) +\theta_{1}\zeta \frac{dF}{d\Psi}|_{\Psi=\Psi_0},
 \label{Ube}
\end{equation}
where $\Psi_0=\kappa(x-\kappa^2 t)$, and $u+\theta_{1}\xi_{1}=\partial_{x}{\cal U}^{b}$, we see that the fermionic solution $\xi_1$ is essentially the derivative with respect to $x$ of $u$, \textit{ie} the particular solution of (\ref{solxiH}). In the conclusion, we discuss the other solution of this second order linear equation.

We easily recover a one super soliton choosing $a_{1}=i$ and $\kappa=1$. In such a case, the Hirota formalism produces the particular travelling wave solution  $u=(\ref{Bsoliton})$.


Now let us exhibit new solutions known as $N$ super soliton solutions \cite{GS,HK,MY,C1,CRG,GS1}.




Indeed, in order to find the $2$ super soliton solution, we first take
\begin{align}
 \tau_{1}&=1+a_{1}e^{\Psi_{1}}+a_{2}e^{\Psi_{2}}+a_{1}a_{2}A_{12}e^{\Psi_{1}+\Psi_{2}},\\
\tau_{2}&=1+b_{1}e^{\Psi_{1}}+b_{2}e^{\Psi_{2}}+b_{1}b_{2}B_{12}e^{\Psi_{1}+\Psi_{2}},
\end{align}
where now $\Psi_{i}=\kappa_{i}x+\omega_{i}t+\theta_{1}\zeta_{i}$ and $a_{i}$ and $b_{i}$ are nonzero even parameters ($i=1,2$). We thus see that we have to take $\kappa_{1}\neq\kappa_{2}$, which has the effect of breaking the symmetry associated with the translation generator ${\cal P}_{t}- c {\cal P}_{x}$. Introducing these expressions in the mKdV bilinear form (\ref{H1}) and (\ref{H2}), the first equation yields the expected dispersion relations:
\begin{equation}
 \omega_{i}+\kappa_{i}^{3}=0, \quad i=1,2.
\end{equation}
The second equation gives the following conditions:
\begin{align}
 b_{i}&=-a_{i}, \ i=1,2,\\
A_{12}&=B_{12}=\left(\dfrac{\kappa_{1}-\kappa_{2}}{\kappa_{1}+\kappa_{2}}\right)^{2}
\end{align}
and a new condition relating the anticommuting variables $\zeta_{1}$ and $\zeta_{2}$ given by ($\kappa_{1}, \kappa_{2}\neq 0$)
\begin{equation}
 \kappa_{1}\zeta_{2}=\kappa_{2}\zeta_{1}.
\end{equation}
Finally, the $\tau$-functions are given by
\begin{align}
 \tau_{1}&=1+a_{1}e^{\Psi_{1}}+a_{2}e^{\Psi_{2}}+a_{1}a_{2}A_{12}e^{\Psi_{1}+\Psi_{2}},\\
\tau_{2}&=1-a_{1}e^{\Psi_{1}}-a_{2}e^{\Psi_{2}}+a_{1}a_{2}A_{12}e^{\Psi_{1}+\Psi_{2}}.
\end{align}
We may enjoy the behavior of the $u(x,t)$ part of the $2$-soliton solution
\begin{align}
 u(x,t)=
-\dfrac{9(10\cosh(\frac{1}{8}(t-4x))+5\cosh(t-x)+8\sinh(\frac{1}{8}(t-4x))+4\sinh(t-x))}{72+41\cosh(\frac{3}{8}(3t-4x))+81\cosh(\frac{1}{8}(7t-4x))+40\sinh(\frac{3}{8}(3t-4x))},
\end{align}
 where the parameters $\kappa_{1}$, $\kappa_{2}$, $a_{1}$, $a_{2}$ are chosen as $\kappa_{1}=2\kappa_{2}=1$ and $a_{1}=a_{2}=i$, so that $\Psi_{1}=(x-t)+\theta_{1}\zeta_{1}$ and $\Psi_{2}=\frac{1}{8}(4x-t)+\theta_{1}\zeta_{2}$. In Figure 3, we give the solution $|u|$ for $t=-10, 0, 10$.

Again, $\xi$ is the derivative of $u$ but we could want to find the other solution of the fermionic equation.

\begin{figure}[!h]
\begin{center}
 \includegraphics[width=4cm]{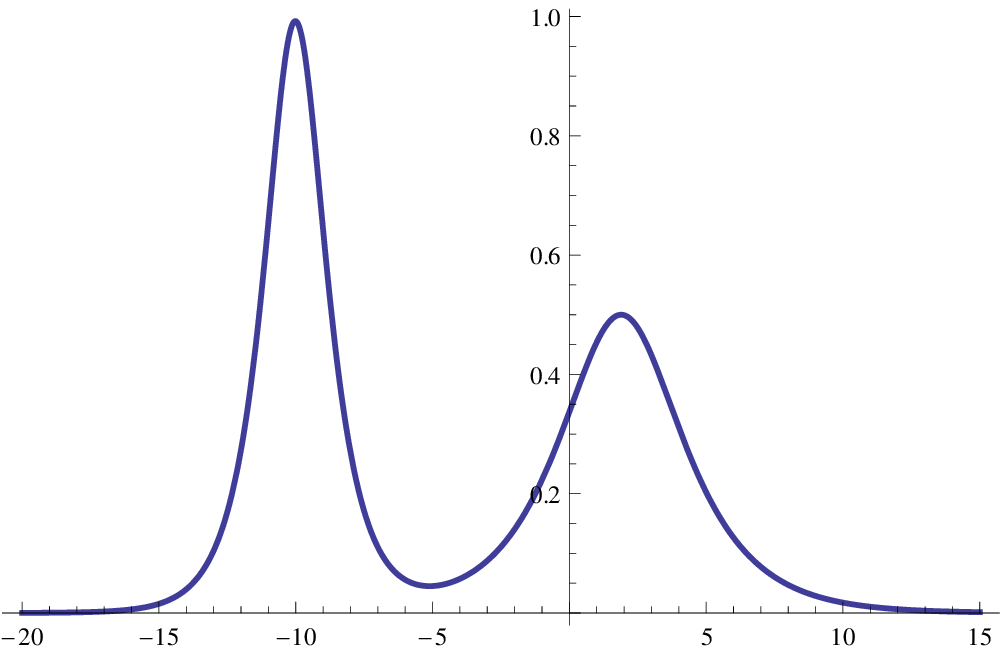}
\includegraphics[width=4cm]{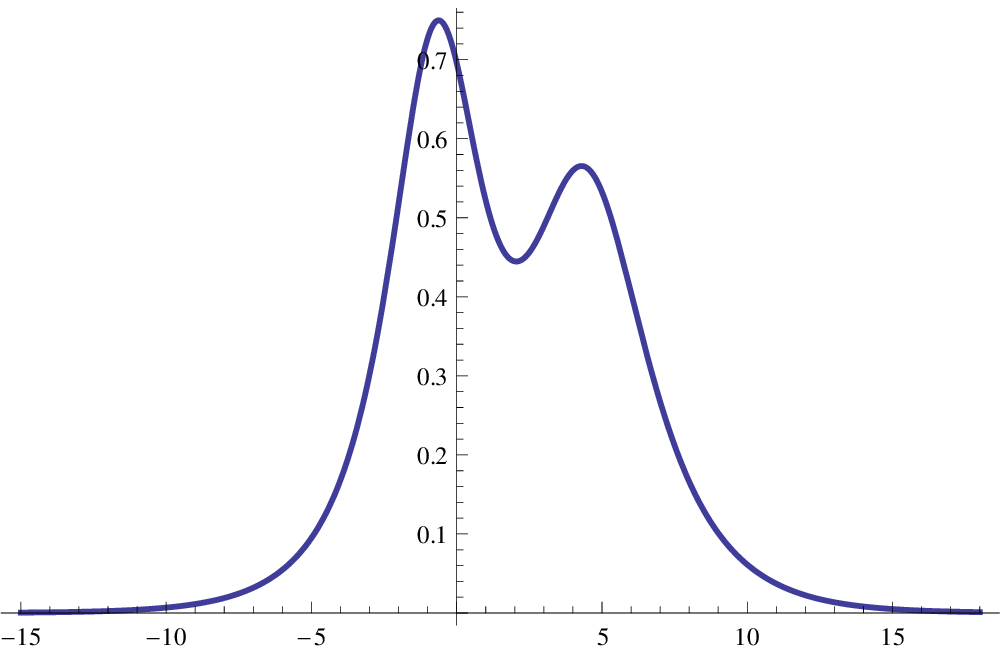}
\includegraphics[width=4cm]{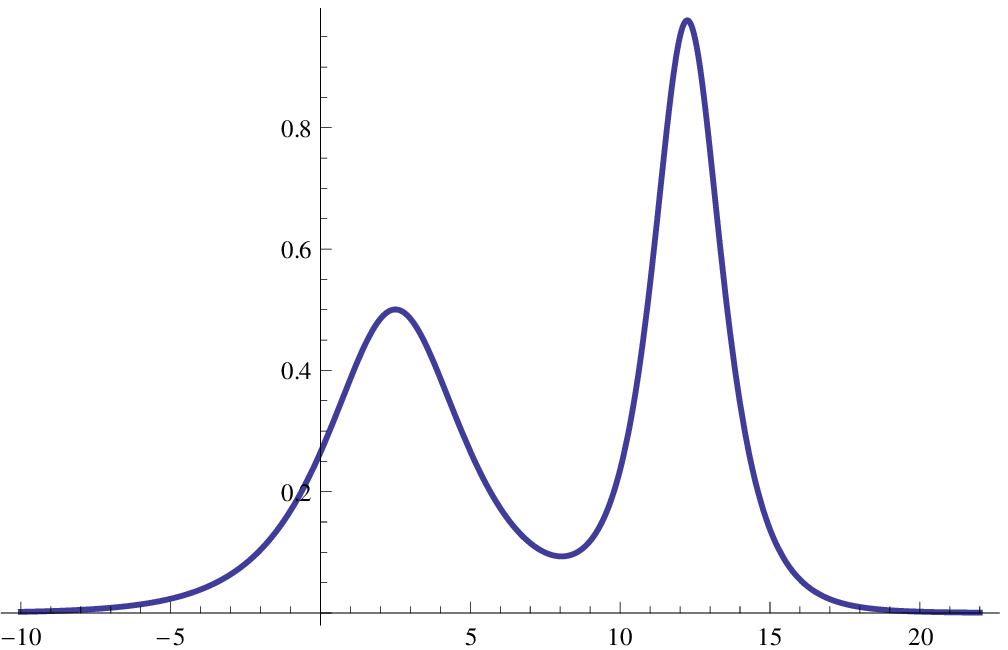}
\caption{The function $|u(x,t)|$ for $t=-10, 0, 10$.}
\end{center}
\end{figure}

 For the three super soliton solution, $\tau_{1}$ and $\tau_{2}$ take the explicit form:
\begin{align}
 \tau_{1}=1+a_{1}e^{\Psi_{1}}+a_{2}e^{\Psi_{2}}+a_{3}e^{\Psi_{3}}+a_{1}a_{2}A_{12}e^{\Psi_{1}+\Psi_{2}}+a_{1}a_{3}A_{13}e^{\Psi_{1}+\Psi_{3}}\notag\\
+a_{2}a_{3}A_{23}e^{\Psi_{2}+\Psi_{3}}+a_{1}a_{2}a_{3}A_{12}A_{13}A_{23}e^{\Psi_{1}+\Psi_{2}+\Psi_{3}},\\
\tau_{2}=1-a_{1}e^{\Psi_{1}}-a_{2}e^{\Psi_{2}}-a_{3}e^{\Psi_{3}}+a_{1}a_{2}A_{12}e^{\Psi_{1}+\Psi_{2}}+a_{1}a_{3}A_{13}e^{\Psi_{1}+\Psi_{3}}\notag\\
+a_{2}a_{3}A_{23}e^{\Psi_{2}+\Psi_{3}}-a_{1}a_{2}a_{3}A_{12}A_{13}A_{23}e^{\Psi_{1}+\Psi_{2}+\Psi_{3}},
\end{align}
where $\Psi_{i}=\kappa_{i}x-\kappa_{i}^{3}t+\theta_{1}\zeta_{i}$ ($i=1,2,3$) and
\begin{align}
 A_{ij}&=\left(\dfrac{\kappa_{i}-\kappa_{j}}{\kappa_{i}+\kappa_{j}}\right)^{2},\\
\kappa_{i}\zeta_{j}&=\kappa_{j}\zeta_{i},
\end{align}
for $i,j=1,2,3$ ($i\neq j$). We may enjoy the behavior of the $u(x,t)$ part of the $3$-soliton solution. In Figure 4, we show again $|u|$ for $t=-15, 0, 15$. The parameters $\kappa_{1}$, $\kappa_{2}$, $\kappa_{3}$, $a_{1}$, $a_{2}$, $a_{3}$ are chosen as $\kappa_{1}=\frac{10}{7}\kappa_{2}=\frac{5}{2}\kappa_{3}=1$ and $a_{1}=a_{2}=a_{3}=i$.

\begin{figure}[!h]
\begin{center}
 \includegraphics[width=4cm]{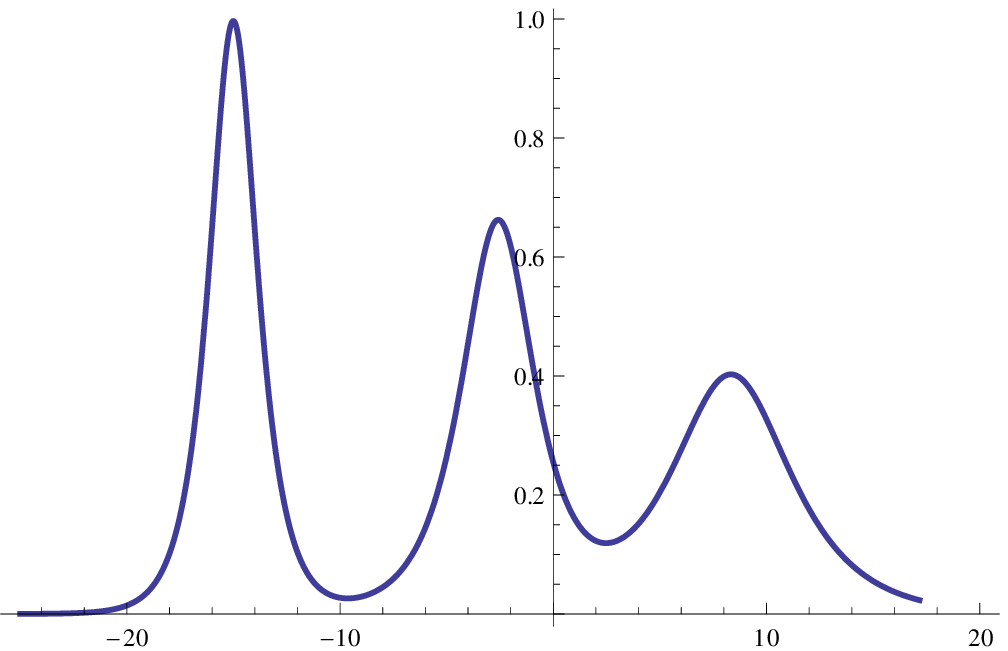}
\includegraphics[width=4cm]{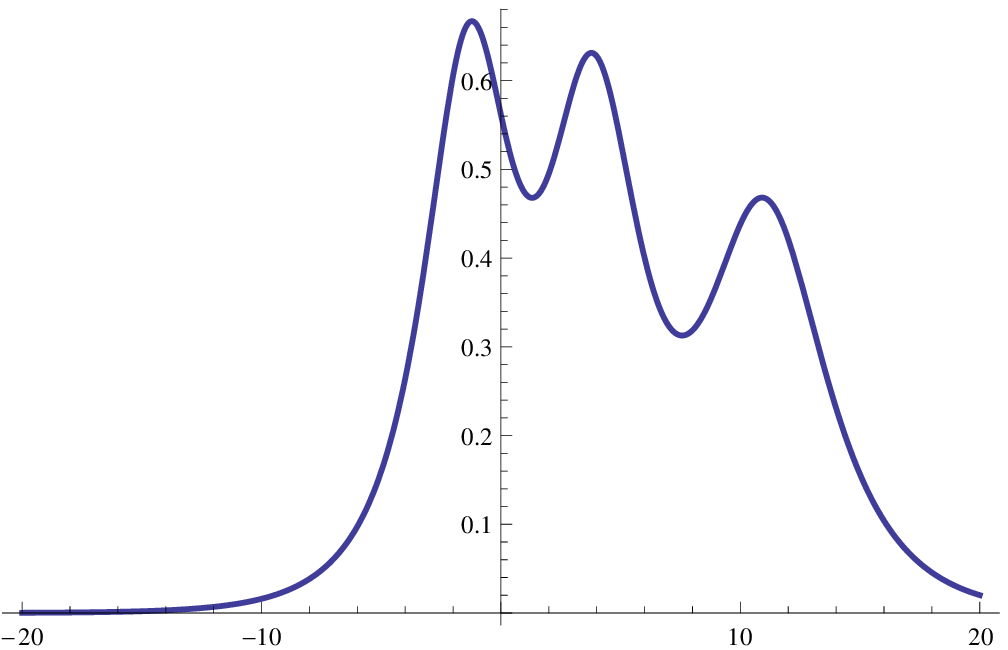}
\includegraphics[width=4cm]{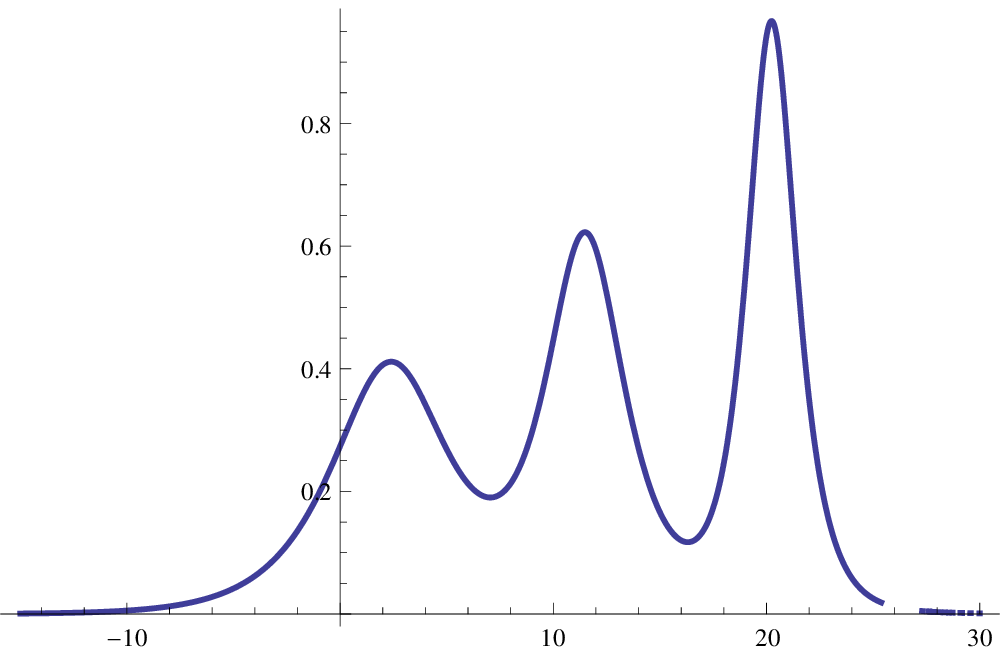}
\caption{The function $|u(x,t)|$ for $t=-15, 0, 15$.}
\end{center}
\end{figure}

 We have thus shown how to construct the $N$ super soliton solution ($N=1,2,3$), using the bilinear form (\ref{H1}) and (\ref{H2}), by giving the explicit forms of the functions $\tau_{1}$ and $\tau_{2}$. The $N$ super soliton solution ($N\geq3$) is easily generalized using the constraints above \cite{AS1}. We are presently working on a \textit{Mathematica} program which generates the $N$ soliton solution by constructing the $\tau$-functions $\tau_{1}$ and $\tau_{2}$ for general $N$.

\subsection{Rational similarity solutions}

The Hirota formalism can also be used to get rational similarity solutions \cite{AS}. We assume now a SUSY generalization where the dependent variables $ \tau_{1}$ and  $\tau_{2}$ are polynomials in the independent variable ${\tilde z}= t^{-\frac13}(x+\theta_{1} \zeta)$ which is similar to (\ref{dilainv}).  To get the SUSY version of the solution (\ref{wsim}), we take
\begin{equation}
 \tau_{1}(x,t;\theta_{1})= \tau_{1}({\tilde z} ,t)= t^{\frac{1}{3}}{\tilde z}, \quad  \tau_{2}(x,t;\theta_{1})= \tau_{2}({\tilde z} ,t)=t({\tilde z}^{3}+12),
 \label{poly1}
\end{equation}
since this choice solves the bilinear system (\ref{H1}) and (\ref{H2}). We thus have the following form 
\begin{equation}
{\cal U}^{b}(x,t;\theta_{1})={\cal U}^{b}({\tilde z} ,t)= i \log\left(t^{\frac{2}{3}}\dfrac{{\tilde z}^3 +12}{{\tilde z}}\right)
 \label{solrs2}
\end{equation}
and, as for the classical case, the invariant solution is thus
\begin{equation}
w_{1}({\tilde z})= t^{\frac{1}{3}}\partial_{x}\ {\cal U}^{b}(x,t;\theta_{1})=i \frac{d}{d{\tilde z}} \log\left(\dfrac{{\tilde z}^3 +12}{{\tilde z}}\right).
\end{equation}
From a Taylor expansion around $\theta_{1}=0$ we get, $z_{0}=xt^{-\frac{1}{3}}$,
\begin{equation}
w_{1}({\tilde z})= w_{1}(z_0)+ t^{-\frac13}\theta_{1} \zeta  \frac{dw_{1}}{d{\tilde z}} |_{{\tilde z}=z_0},
\end{equation}
and since $u^b=u+\theta_1 \xi_1=\partial_x {\cal U}^b=t^{-\frac13} w_{1}({\tilde z})$,
we get 
$u(x,t)=t^{-\frac13}  w_{1}(z_0)$ as in (\ref{identdila}) and $\xi_1=\zeta u_x$, as expected.
The generalization to the infinite set of solutions given in \cite{K,C} is direct and we get
\begin{equation}
{\cal U}^{b}_{n}({\tilde z},t)=i\log\left(t^{\frac{n+1}{3}}\dfrac{Q_{n+1}({\tilde z})}{Q_{n}({\tilde z})}\right),\label{solvb}
\end{equation}
where the functions $Q_{n}({\tilde z})$ are the Yablonskii-Vorob'ev polynomials define in Appendix 1. The link with the Hirota formalism is obtained by letting $ \tau_{1}$ and  $\tau_{2}$ be polynomials in the independent variable $\tilde z$ and the time variable $t$. In fact, we take the following series
\begin{equation}
 \tau_{1,n}({\tilde z},t)=t^{\frac{n(n+1)}{6}}Q_{n}({\tilde z}), \ \ \tau_{2,n}({\tilde z},t)=t^{\frac{(n+1)(n+2)}{6}}Q_{n+1}({\tilde z})
\end{equation}
which lead to (\ref{solvb}). For example, we see that if $n=1$, we recover the expression (\ref{poly1}).

The corresponding invariant solutions $w({\tilde z})$ of (\ref{redw2}) are easily obtained by derivation
\begin{equation}
 w_{n}({\tilde z})=i\frac{d}{d{\tilde z}}\log\left(\dfrac{Q_{n+1}({\tilde z})}{Q_{n}({\tilde z})}\right),\quad c_{2}=c_{2;n}=i\frac{(n+1)}{3}.
\end{equation}
where we see that the integration constant $c_2$ is different for each solution, \textit{ie} $c_{2}=c_{2;n}$. We also have the following identity
\begin{equation}
 {\cal U}^{b}_n({\tilde z},t)= c_{2;n}\log t+\int w_{n}({\tilde z})d{\tilde z}.
\end{equation}
We may want to give the corresponding solutions for $u_{n}(x,t)$, in terms of the independent variables $x$ and $t$. For example, from $w_{2}({\tilde z})$, we find
\begin{equation}
 u_{2}(x,t)=\dfrac{3it^{\frac{1}{3}}x^{2}(x^{6}+24tx^{3}+1440t^{2})}{x^{9}+72tx^{6}-8640t^{3}},
\end{equation}
and, using $w_{3}({\tilde z})$, we get
\begin{align}
 u_{3}(x,t)= \dfrac{4it^{\frac{1}{3}}(x^{15}+150tx^{12}+9000t^{2}x^{9}-604800t^{3}x^{6}-9072000t^{4}x^{3}-54432000t^{5})}{x(x^{6}+60tx^{3}-720t^{2})(x^{9}+180tx^{6}+302400t^{3})}.
\end{align}


\section {Conclusions \label{sec:conclusions}}

The resolution of the $\mathcal{N}=2$ SUSY KdV equation for $a=-2$ has been revised with the aim to show some links between the symmetry reduction method and the Hirota formalism. In the first case, we have fully used subalgebras and invariants including bosonic and fermionic dependent and independent variables to get  travelling wave solutions and also rational similarity solutions.
In the second case, we have been able, for the first time, to extend the Hirota formalism to the $\mathcal{N}=2$ SUSY case and to produce solutions of the SUSY KdV equation which are similar to the preceding ones but not identical. Indeed, the assumption made has produced a reduction of our equation to a $\mathcal{N}=1$ SUSY one for which the bilinearization is known. 
From this last formalism, we recover the large set of solitonic solutions ($N$ solitons) that has already been found in the classical case and in the  $\mathcal{N}=1$ SUSY case. More interestingly, similarity rational solutions have been generalized in the SUSY context and we have shown that they appear as well when we consider the Hirota formalism. We have shown in both cases that $\xi_{1}=\zeta u_{x}$ solves the fermionic equation (\ref{solxiH}) for a solution $u$ of (\ref{soluH}). The Hirota formalism has produced only one independent solution of the second order linear differential equation (\ref{solxiH}). Using classical resolution technics, one can easily retrieve the second linearly independent solution of the homogeneous fermionic equation for which the Hirota formalism has not yet been adapted to find such solutions. Indeed, for $y_{0}=x-t$ and $u(y_{0})={\text{sech}}\ y_{0}$, one gets 
\begin{equation}
 \xi_{1}(y_{0})=\zeta{\text{sech}}\ y_{0}(-5+\cosh2y_{0}+6y_{0}\tanh y_{0}).
\end{equation}
For $z_{0}=xt^{-\frac{1}{3}}$, $u(x,t)=t^{-\frac{1}{3}}w_{0}(z_{0})$ and $\xi_{1}=t^{-\frac{1}{3}}\phi_{0}(z_{0})$ we get
\begin{equation}
 \phi_{0}(z_{0})=\zeta\sqrt{z_{0}}I_{\frac{5}{3}}\left(\frac{2}{3\sqrt{3}}z_{0}^{\frac{3}{2}}\right),
\end{equation}
which is a solution of the second order linear differential equation,
\begin{equation}
 \phi^{\prime\prime}+\left(6w^{2}-\dfrac{1}{3}z_{0}\right)\phi=0.
\end{equation}

\section{Acknowledgments}

L. Delisle acknowledge the support of a FQRNT doctoral research scholarship. V. Hussin acknowledge the support of research grants from NSERC of Canada. 

\section{Appendix I: On rational similarity solutions of mKdV}
From the approach of Clarkson \cite{C} adapted to our context, we define the following recurrence relation for the construction of the Yablonskii-Vorob'ev polynomials,
\begin{equation}
 3^{\frac{1}{3}}Q_{n+1}Q_{n-1}=zQ_{n}^{2}-12\left(Q_{n}Q_{n;zz}-Q_{n;z}^{2}\right),
\end{equation}
with $Q_{0}(z)=3^{-\frac{1}{3}}$ and $Q_{1}(z)=z$. Here are some examples,
\begin{align}
 Q_{2}(z)&=z^{3}+12,\\
Q_{3}(z)&=3^{-\frac{1}{3}}\left(z^{6}+60z^{3}-720\right),\\
Q_{4}(z)&=\frac{1}{3}z\left(z^{9}+180z^{6}+302400\right).
\end{align}
Using the above polynomials, we can thus construct a series of rational similarity solutions of equation (\ref{redw2}) as shown in \cite{C}, with a slight change of variables:
\begin{equation}
 w_{n}^{\pm}(z)=\pm i \dfrac{d}{dz}\log\left(\dfrac{Q_{n}(z)}{Q_{n+1}(z)}\right),
\end{equation}
where the constant of integration $c_{2}$ is given by
\begin{equation}
 c_{2;n}^{\pm}=\mp i \frac{(n+1)}{3}.
\end{equation}
For example, the first solutions $w_{n}^{\pm}(z)$ ($w_{n}^{-}=-w_{n}^{+}$) of the reduce equation (\ref{redw2}) are given as:
\begin{align}
w_{0}^{-}(z)&=\frac{i}{z} , \\
w_{1}^{-}(z)&=2i \frac{z^3-6}{z(z^3+12)},
\label{wsim1}\\
w_{2}^{-}(z)&=3i \frac{z^{2}(z^{6}+24z^{3}+1440)}{z^{9}+72z^{6}-8640}, \\
w_{3}^{-}(z)&=4i \frac{z^{15}+150z^{12}+9000z^{9}-604800z^{6}-9072000z^{3}-54432000}{z(z^{6}+60z^{3}-720)(z^{9}+180z^{6}+302400)}.
\end{align}
Interestingly, the fermionic equation (\ref{redphi2}) may be solved using some properties of the solutions $w_{n}^{\pm}$. Indeed, let us define
\begin{equation}
 \mathcal{W}_{n}^{\pm}=-\frac{1}{3}z+6(w_{n}^{\pm})^{2}+6iw_{n;z}^{\pm},
\label{coefphi}
\end{equation}
so that equation (\ref{redphi2}) with $k_{2}=0$ becomes
\begin{equation}
 \phi_{n;zz}^{\pm}+\mathcal{W}_{n}^{\pm}\phi_{n}^{\pm}=0.
\end{equation}
Depending on the choice of $w_{n}^{+}$ or $w_{n}^{-}$, expression (\ref{coefphi}) takes the form:
\begin{align}
 \mathcal{W}_{n}^{+}&=\frac{2}{3}z+i \dfrac{2n+3}{w_{n}^{+}+w_{n+1}^{+}},\\
\mathcal{W}_{n}^{-}&=-\frac{4}{3}z+12(w_{n}^{-})^{2}+i \dfrac{2n+3}{w_{n}^{-}+w_{n+1}^{-}}.
\end{align}
One can see that we have the following property,
\begin{equation}
\mathcal{W}_{n+1}^{-}= \mathcal{W}_{n}^{+},\label{propW}
\end{equation}
for $n\geq0$ and $\mathcal{W}_{0}^{-}=-\dfrac{z}{3}$. We get explicitly,
\begin{align}
 \mathcal{W}_{1}^{-}&=-\dfrac{z^{3}+36}{3z^{2}},\\
\mathcal{W}_{2}^{-}&=-\dfrac{z(z^{6}+132z^{3}-2448)}{3(z^{3}+12)^{2}},\\
\mathcal{W}_{3}^{-}&=-\dfrac{z(z^{12}+336z^{9}+2160z^{6}+1080000z^{3}+9849600)}{3(z^{6}+60z^{3}-720)^{2}}.
\end{align}
From (\ref{propW}), we deduce that $\phi_{n+1}^{-}=\phi_{n}^{+}$ for $n\geq0$. We also get
\begin{equation}
 \phi_{0;zz}^{-}+\mathcal{W}_{0}^{-}\phi_{0}^{-}=\phi_{0;zz}^{-}-\dfrac{z}{3}\phi_{0}^{-}=0,
\end{equation}
has a solution expressible in terms of Airy functions. The particular solution $\phi=$(\ref{solphi}), solves in fact,
\begin{equation}
 \phi_{1;zz}^{-}+\mathcal{W}_{1}^{-}\phi_{1}^{-}=0.
\end{equation}

\section{Appendix II: Bilinear identities}

We present here some identities that are useful to simplify some calculations. In fact, taking $\Psi_{i}=\kappa_{i}x+\omega_{i}t+\theta\zeta_{i}$, we get the following identities:
\begin{align*}
 \mathcal{S}\mathcal{D}_{x}^{n}(e^{\Psi_{1}}\cdotp e^{\Psi_{2}}) &=(\zeta_{1}-\zeta_{2}+\theta(\kappa_{1}-\kappa_{2})) (\kappa_{1}-\kappa_{2})^{n}e^{\Psi_{1}+\Psi_{2}},\\
\mathcal{P}(\mathcal{D}_{x},\mathcal{D}_{t})(e^{\Psi_{1}}\cdotp 1)&=\mathcal{P}(\partial_{x},\partial_{t})e^{\Psi_{1}},
\end{align*}
where $\mathcal{P}$ is a polynomial. From these identities, we deduce useful results:
\begin{align*}
 \mathcal{S}\mathcal{D}_{x}^{2k}(e^{\Psi_{1}}\cdotp e^{\Psi_{2}})&=-\mathcal{S}\mathcal{D}_{x}^{2k}(e^{\Psi_{2}}\cdotp e^{\Psi_{1}}),\\
\mathcal{S}\mathcal{D}_{x}^{2k+1}(e^{\Psi_{1}}\cdotp e^{\Psi_{2}})&=\mathcal{S}\mathcal{D}_{x}^{2k+1}(e^{\Psi_{2}}\cdotp e^{\Psi_{1}}),\\
\mathcal{S}\mathcal{D}_{x}^{n}(e^{\Psi_{1}}\cdotp 1)&=(\zeta_{1}+\theta\kappa_{1})\kappa_{1}^{n}e^{\Psi_{1}},\\
\mathcal{S}\mathcal{D}_{x}^{n}(e^{\Psi_{1}}\cdotp e^{\Psi_{1}}) &=0.
\end{align*}
For example, if $\tau_{1}=1+\delta e^{\Psi}$ and $\tau_{2}=1-\delta e^{\Psi}$ with $\Psi$ given as above, we get
\begin{eqnarray*}
 (\mathcal{D}_{t}+\mathcal{D}_{x}^{3})(\tau_{1}\cdotp\tau_{2})&=&(\mathcal{D}_{t}+\mathcal{D}_{x}^{3})(1\cdotp 1+\delta e^{\Psi}\cdotp 1-\delta 1 \cdotp e^{\Psi}-\delta^{2}e^{\Psi}\cdotp e^{\Psi})\\
&=&2\delta (\omega+\kappa^{3})e^{\Psi}.
\end{eqnarray*}


\clearpage

\section*{References}

\begin{thebibliography}{99}

\bibitem{LM}
P. Labelle, P. Mathieu (1991) A new supersymmetric
Korteweg\/--\/de Vries equation, {\it J.~Math.\ Phys.}{\bf 32},
923--927.

\bibitem{AHW}
M. A. Ayari, V. Hussin, and P. Winternitz (1999)
Group invariant solutions for the $N=2$ super Korteweg\/--\/de Vries
equation, {\it J.~Math.\ Phys.} {\bf40}, 1951--1965.

\bibitem{IMM}
A. Ibort, L. Mart\'\i nez Alonso, and E. Medina Reus (1996)
Explicit solutions of supersymmetric KP hierarchies: supersolitons and
solitinos, {\it J.~Math.\ Phys.} {\bf 37}, 6157--6172.

\bibitem{GS}
S. Ghosh, D. Sarma (2001) Soliton solutions for the $N=2$
supersymmetric KdV equation, {\it Phys.\ Lett.~B} {\bf 522}, 189--193.

\bibitem{HK}
V. Hussin and A. V. Kiselev (2009) 
Virtual Hirota' s multi-soliton solutions of N=2 supersymmetric Korteweg\/--\/de Vries equations, 
{\it Theor. Math. Phys.} {\bf159}, 832--840.

\bibitem{ARS}
S. Andrea, A. Restuccia and A. Sotomayor (2001) An operator valued extension of the super KdV equations. {\it J.~Math.\ Phys.} {\bf 42}, 2625--2634.

\bibitem{K} 
Y. Kametaka (1983) On rational similarity solutions of KdV and mKdV equations, {\it Proc. Japan. Acad.}, {\bf 59},	407--409.

\bibitem{DJ}
P. G. Drazin and R. S. Johnson (1989) Solitons: an introduction, {\it Cambridge University Press}.

\bibitem{FOU}
S. Fukutani, K. Okamoto and H. Umemura (2000)
Special polynomials and the Hirota bilinear relations of the second and the fourth Painlev\'e equations,
{\it Nagoya Math. J.} {\bf159}, 179--200.

\bibitem{AS}
M. J. Ablowitz and J. Satsuma (1978)
Solitons and rational solutions of nonlinear evolution equations,
{\it J.~Math.\ Phys.} {\bf 19}, 2180--2186.

\bibitem{C}
P.A. Clarkson (2003) Remarks on the Yablonskii-Vorob'ev polynomials,{\it Physics Lett.~A} {\bf 319}, 137--144.

\bibitem{MY}
I.N. Mc Arthur and C. M. Yung (1993) Hirota bilinear form for the super-KdV hierarchy, {\it Modern Physics Lett.~A} {\bf 8}, 1739--1745.

\bibitem{C1}
A. S. Carstea (2000) Extension of the bilinear formalism to supersymmetric KdV\/-\/type equations, {\it Nonlinearity} {\bf 13}, 1645--1656.

\bibitem{CRG}
A. S. Carstea, A. Ramini and B. Grammaticos (2001) Constructing the soliton solutions of the $N=1$ supersymmetric KdV hierarchy, {\it Nonlinearity} {\bf 14}, 1419--1423.

\bibitem{GS1}
S. Ghosh, D. Sarma (2003) Bilinearization of $N=1$ supersymmetric
modified KdV equations, {\it Nonlinearity} {\bf16}, 411--418.

\bibitem{ZLW} 
M.-X. Zhang, Q. P. Liu, Y.-L. Shen and K. Wu (2008)
Bilinear approach to N = 2 supersymmetric KdV equations, {\it Science in China Series A: Mathematics}, {\bf52}, 1973--1981.

\bibitem{AS1}
M. J. Ablowitz and H. Segur (1981) Solitons and the Inverse Scattering Transform, {\it SIAM}.


\end{thebibliography}
\end{document}